\documentclass{elsart}
\usepackage{amssymb}
\usepackage{amsmath}

\newcommand{\bea}{\begin{eqnarray}}
\newcommand{\eea}{\end{eqnarray}}
\newcommand{\be}{\begin{equation}}
\newcommand{\ee}{\end{equation}}

\newcommand{\rt}[1]{{}}

\begin{document}

\begin{frontmatter}
\title{Phantom field fluctuation induced Higgs effect}

\author[label1,label2]{A. Patk{\'o}s}
\address[label1]{Department of Atomic Physics, E{\"o}tv{\"o}s University,
H-1117 Budapest, Hungary}
\address[label2]{Research Group for Statistical Physics of the
Hungarian Academy of Sciences, \\H-1117 Budapest, Hungary}
\ead{patkos@ludens.elte.hu}

\author{Zs. Sz{\'e}p}
\address{Research Institute for Solid State Physics and Optics of the
  Hungarian Academy of Sciences, H-1525 Budapest, Hungary}
\ead{szepzs@achilles.elte.hu}

\begin{abstract}
Symmetry breaking solutions are investigated in the $N\rightarrow \infty$ 
limit for the ground state of a system consisting of a Lorentz-scalar, 
$N$ component ``phantom'' field and an $O(N)$ singlet. The most general
form of $O(N)\times Z_2$ invariant quartic interaction is
considered. The non-perturbatively renormalised solution demonstrates
the possibility for $Z_2$ symmetry breaking induced by phantom
fluctuations. It becomes also evident that the strength of the
``internal'' dynamics of the N-component field tunes away the ratio of
the Higgs condensate and the Higgs mass from its perturbative (nearly
tree-level) expression. 
\end{abstract}


\end{frontmatter}

\section{Introduction}
Scalar fields are proposed at present with renewed intensity as
minimal, phenomenologically motivated complements to the Standard
Model. The simplest mechanism for electroweak symmetry breaking is
realised by introducing a renormalisable biquadratic coupling of the
Higgs field to a ``hidden scalar'' (the phantom field) which acquires
nonzero vacuum expectation value \cite{wilczek05,calmet06}.
Postulation of two extra scalars with appropriate dynamics might
provide the minimal framework for the particle physics interpretation
of the cosmologically well-established facts of missing luminous mass
and of inflationary density fluctuations \cite{murayama05,bij06}. A
scalar inflaton coupled to sterile right-handed neutrinos offers
another cosmologically motivated 
minimal extension of the standard model, which accounts
also for neutrino oscillation phenomena
 and the fermion-asymmetry of the universe \cite{shapo06}.

A general background to this evolution is provided by low energy field
theories constructed on string vacua showing maximal compatibility
with the particle content of the Standard Model. Appearance of extra
scalars is characteristic for all these constructions. A consequence
of possible physical significance is that the natural cut-off scale to
be applied for the top quark contribution to the Higgs mass would be
raised considerably \cite{hall05}. Consequences of the existence of a
phantom sector on the precision electroweak tests of the
Standard Model are being systematically explored \cite{Candeno}. 
The history of such investigations goes back to the 80's \cite{hill87}.

The goal of this letter is to illustrate the substantial influence of
an extended scalar sector on the symmetry breaking vacuum features
through the results of an exact, non-perturbatively renormalised large
$N$ analysis. We propose the following embedding of the Higgs sector
of the Standard Model into a wider set of elementary scalar
fields. The gauged self-interacting scalar $SU(2)$ doublet is coupled
also to a set of scalar fields which form an $N$-component vector
under a hidden internal $O(N)$ symmetry. The coupling preserves the
$SU(2)$ symmetry of the standard gauge model. After fixing 
3 components of the complex Higgs doublet to zero
a $Z_2$ invariance of the Higgs field is left. The phantom
$O(N)$ vector field has some extra $O(N)$ invariant non-linear
interaction on its own. With appropriately chosen $N$-scaling of the
quartic couplings the renormalised $N=\infty$ solutions will be
presented. We shall demonstrate that the non-perturbative
renormalisation of the solution leads to a non-trivial $Z(2)$ symmetry
breaking Higgs condensate in a large range of the renormalised
couplings of the model even if no negative renormalised squared mass
is introduced.  It will be shown that also essential deviations might
appear in the relation of the Higgs mass to the strength of the Higgs
condensate, relative to the leading order large $N$ estimate derived
in the $O(N)$ symmetric linear sigma model.

The present model is a generalisation of the $O(N+M)$ symmetric model
studied by Chivukula {\it et al.} in 1991-1993 \cite{chivukula91}. 
Another extension of the Higgs sector was investigated in \cite{ignatiev00}, 
where consequences of the mirror world assumption were studied.
Here the general model will be solved by taking the large $N$ limit of 
the renormalised Dyson-Schwinger equations for the exact $n$-point
functions. The renormalisation process will follow an approach
advocated recently in the context of 2PI approximate solution of
scalar field theories \cite{reinosa04}, but we emphasise that it
represents a rather non-trivial implementation of the strategy of
nonperturbative renormalisation.

\section{Leading order formal solution}

The Lagrangian to be solved at $N=\infty$ 
describes the most general renormalisable $SU(2)\times
O(N)$ symmetric interaction of an $N$-component (and SU(2) singlet) field
$\psi_i, ~i=1,..,N$ with a gauged $SU(2)$ doublet (and $O(N)$ singlet) $\Phi$:
\begin{eqnarray}
L[\psi_i,\phi]&=&\frac{1}{2}(D_\mu\Phi)^\dagger
D^\mu\Phi+\frac{1}{2}(\partial_\mu\psi_i)^2-\frac{1}{2}m_2^2\psi_i^2-
\frac{1}{2}m_3^2\Phi^\dagger\Phi
\nonumber\\
&&-\frac{\lambda_1}{24N}(\psi_i^2)^2-\frac{\lambda_2}{24N}
(\Phi^\dagger\Phi)^2-\frac{\lambda_3}{12N}\psi_i^2\Phi^\dagger\Phi.
\end{eqnarray}
The present parametrisation is a generalisation of that used in
\cite{chivukula91}, where only a unique quartic coupling $\sim
(\psi_i^2+\Phi^\dagger\Phi)^2$ 
was considered.

We shall investigate the usual symmetry breaking pattern, where the
lower real part of the complex $SU(2)$ doublet gets a nonzero classical
value $\sqrt{N}v$, and the other three components are gauge transformed
away (unitary gauge). The shift in the Higgs field then breaks the
$Z_2$ symmetry related to the remaining reflection symmetry 
of the Higgs field. When gauge fields are set to zero
the resulting shifted Lagrangian density is the following:
\begin{eqnarray}
L_\textnormal{free}[\psi_i,\sigma]&=&\frac{1}{2}(\partial_\mu\sigma)^2+
\frac{1}{2}
(\partial_\mu\psi_i)^2-\frac{1}{2}m_2^2\psi_i^2-\frac{1}{2}m_3^2
(\sigma^2+2\sigma\sqrt{N}v+Nv^2),\nonumber\\
L_\textnormal{int}[\psi_i,\sigma]&=&-\frac{\lambda_1}{24N}(\psi_i^2)^2-\frac{\lambda_2}
{24 N}(\sigma^2+2\sigma\sqrt{N}v+Nv^2)^2\nonumber\\
&&-\frac{\lambda_3}{12N}\psi_i^2
(\sigma^2+2\sigma\sqrt{N}v+Nv^2).
\label{auxL}
\end{eqnarray}

The large $N$ solutions will be constructed with help of the Dyson-Schwinger
(DS) equations which are generated by the master equation
\begin{equation}
\frac{\delta\Gamma}{\delta\varphi_U}=\frac{\delta S_{cl}}{\delta\varphi_U}
\left[\varphi_A+G_{AB}\frac{\delta}{\delta\varphi_B}\right],
\end{equation}
and its functional derivatives \cite{Rivers}. 
The right hand side is evaluated by
replacing in the expression of the functional derivative of the
classical action the field $\varphi_U$ by
$\varphi_U+G_{UV}\delta /\delta\varphi_V$ and the whole expression is
applied to the unity in the field space. $\Gamma$ is the effective
action, $\varphi_U$ is a generic field variable, $G_{AB}$ is the
two-point function of the generic fields $\varphi_A$ and $\varphi_B$
(deWitt's convention is used for the index $A,B,...$). Note, that one
should include into the intermediate steps of the derivation all
possible ($A,B$) pairs of indices for $G_{AB}$, the corresponding
external currents are set to zero only at the very end.

Below we give the unrenormalised equations which determine
the vacuum condensate $v$, the propagators 
$G_{\sigma\sigma}(x,y)=:G_{\sigma}(x,y)$, 
$G_{\psi_m\psi_n}(x,y)=:\delta_{nm}G_{\psi}(x,y)$ and the relevant
3-point function $\Gamma_{\psi_n\psi_m\sigma}(x,y,z)=
\delta_{nm}\Gamma_{\psi\psi\sigma}(x,y,z)$. 
The LO ($N=\infty$) equation for the vacuum expectation value $v$
looks rather simple:
\begin{equation}
-\frac{\delta\Gamma}{\delta\sigma}_{|\varphi_U=0}=\sqrt{N}v
\left[ m_3^2+\frac{\lambda_2}{6}v^2+\frac{\lambda_3}{6}T_\psi
\right]=0,\qquad
T_\psi=\int_kG_{\psi}(k).
\label{extrema}
\end{equation}
The LO $\psi$ propagator is obtained as 
$\delta^2\Gamma/\delta\psi\delta\psi=:iG^{-1}_{\psi}(p)$ in the
following form
\begin{equation}
iG^{-1}_{\psi}(p)=p^2-m_2^2-\frac{\lambda_3}{6}v^2-\frac{\lambda_1}{6}T_\psi.
\label{psiM}
\end{equation}
Since the tadpole is momentum independent the propagator can be
parametrised as $G_\psi(p)=i/(p^2-m^2_\psi)$ where $m^2_\psi$ is 
determined by the gap-equation
\be
m_\psi^2=m_2^2+\frac{\lambda_3}{6}v^2+\frac{\lambda_1}{6}T_\psi.
\label{Eq:gap-equation}
\ee
Finally, the closed system determining the propagator $G_{\sigma}$ and
the 3-point function $\Gamma_{\psi\psi\sigma}$ is derived by taking
the corresponding second and 
third functional derivatives of the effective action:
\begin{eqnarray}
iG_{\sigma}^{-1}(p)&=&p^2-m_3^2-\frac{\lambda_2}{2}v^2-
\frac{\lambda_3}{6}T_\psi\nonumber
\\&&-i\sqrt{N}v\frac{\lambda_3}{6}\int_k
G_{\psi}(p-k)G_{\psi}(k)
\Gamma_{\psi\psi\sigma}(p-k,k,-p),\nonumber\\
\Gamma_{\psi\psi\sigma}(p,q,-p-q)&=&-\frac{\lambda_3 v}{3\sqrt{N}}
\nonumber\\
&&-i\frac{\lambda_1}{6}\int_k G_{\psi}(p+q-k)G_{\psi}(k)
\Gamma_{\psi\psi\sigma}(p+q-k,k,-p-q).
\label{MsigmaM}
\end{eqnarray}
We note that 
$\Gamma_{\psi\psi\sigma}$ was calculated from the master equation of 
$\delta\Gamma/\delta\psi$. 
In the last equation the expression of the right hand side
shows that $\Gamma_{\psi\psi\sigma}(p,q,-p-q)$ on the left depends only on the
momentum of $\sigma$. From this it promptly follows 
for the 3-point function that
\begin{equation}
\Gamma_{\psi\psi\sigma}(p,q,-p-q)=-\frac{\lambda_3v}{3\sqrt{N}}\frac{1}
{1-\frac{\lambda_1}{6}I_\psi(p+q)},\qquad
I_\psi(p)=-i\int_kG_{\psi}(p-k)G_{\psi}(k),
\label{Gamma}
\end{equation}
which leads  (by exploiting Eq.(\ref{extrema})) to
\begin{equation}
iG_{\sigma}^{-1}(p)=p^2-\frac{\lambda_2}{3}v^2-\frac{\lambda_3^2}
{18}v^2\frac{I_\psi(p)}{1-\frac{\lambda_1}{6}I_\psi(p)}.
\label{sigmaMM}
\end{equation}

The renormalisation algorithm of the equations (\ref{extrema}),
(\ref{psiM}) and (\ref{MsigmaM}) is rather non-trivial.  The result
can be stated, however, very simply: i) one should replace in these
equations the bare couplings $m_i^2, \lambda_i$ by the corresponding
renormalised ones; ii) the quadratically divergent tadpole integral
$T_\psi$ and the logarithmically divergent $I_\psi$ should be replaced
by the finite parts $T_{\psi,F}, I_{\psi,F}$ arising after subtracting
the divergent pieces.
It is worth noting, that by the renormalisation of 
$\lambda_i$ the 3-point function $\Gamma_{\psi\psi\sigma}$ is actually a finite
asymptotically decreasing function of the $\sigma$-momentum. As a consequence
no divergence proportional to $p^2$ appears in $G_\sigma^{-1}$, therefore 
there is no need for infinite field renormalisation.

The derivation of the non-trivial relation of the renormalised 
couplings to the unrenormalised ones will be outlined in the next section.
We  apply cut-off regularisation to the divergent integrals 
$T_\psi(M)$ and $I_\psi(p,M)$, which are evaluated after Euclidean 
continuation of the variables:
\begin{eqnarray}
T_\psi(M)&\equiv& \int_k\frac{i}{k^2-M^2}=\frac{\Lambda^2}{16\pi^2}-\frac{M^2}
{16\pi^2}\ln\frac{e\Lambda^2}{M_0^2}+T_{\psi,F}(M),\nonumber\\
I_\psi(p,M)&\equiv& \int_k\frac{i}{(k^2-M^2)((p-k)^2-M^2)}=-\frac{1}{16\pi^2}
\ln\frac{e\Lambda^2}{M_0^2}+I_{\psi,F}(p,M),
\end{eqnarray}
where the scale $M_0$ is introduced to eliminate 
the $M$-dependence of the cut-off dependent terms. 
The renormalisation will be achieved
by compensating the cut-off dependence explicitly written out above by 
appropriately chosen coupling counterterms.

\section{Renormalisation}

The ultraviolet renormalisation of the equations (\ref{extrema}),
(\ref{psiM}) and (\ref{MsigmaM}) requires a fresh analysis of the
counter-term resummation, since the unrenormalised equations do not
suggest any transparent formula similar to the simpler case of 
the $O(N)$ model (see {\it e.g.} \cite{Moshe83}).

Below we shall apply the iterative renormalisation of \cite{reinosa04}
to our case. The renormalisation of the equation of state and pion
propagator (gap-equation) is the usual superdaisy renormalisation
described in details in \cite{reinosa04}.  The renormalisation of the
system (\ref{MsigmaM}) is a little bit more involved. First one
rewrites the previous equations by making explicit the
counterterms. Since the derivation of the DS equations starts from the
classical action, there are explicit Lagrangian parameters present
which we have to split as $m_i^2=m_{i,R}^2+\delta m_i^2,
\lambda_i=\lambda_{i,R}+\delta\lambda_i$. But the effect of the
counterterm does not reduce merely to this. Their consistent
application implies that the $3$-point function entering the DS
equation of the sigma propagator is the renormalised one.  This
statement can be easily checked when using in the equation
of the sigma propagator the iterative solution of the equation of the
renormalised 3-point function: 
\bea
\Gamma_{\psi\psi\sigma}(p,q,-p-q)&=&-
\frac{\lambda_{3,R}+\delta\lambda_3}{3\sqrt{N}}v\\ \nonumber
&&-i\frac{\lambda_{1,R}+\delta\lambda_1}{6}\int_k
G_{\psi}(p+q-k)G_{\psi}(k) \Gamma_{\psi\psi\sigma}(p+q-k,k,-p-q).
\eea 
At a given order of the iteration we recover the chain of bubble
diagrams including the counterterm diagrams of the usual
perturbation theory at the corresponding order of the loop expansion.

Next we subtract Eq.~(\ref{extrema}) from the first equation of
(\ref{MsigmaM}) to obtain for the renormalised
$iG_{\sigma}^{-1}$ the expression
\begin{equation}
iG_{\sigma}^{-1}(p)=p^2-\frac{\lambda_{2,R}+\delta\lambda_2}{3}v^2-i
\sqrt{N}v\frac{\lambda_{3,R}+\delta\lambda_3}{6}\int_kG_{\psi}(p-k)
G_{\psi}(k)\Gamma_{\psi\psi\sigma}(p-k,k,-p).
\end{equation}
The iterative renormalisation consists of assuming
infinite series expansion for the counterterms and determining them by
requiring the cancellation of the divergencies generated at each
iteration step of the solution of the coupled 
equations of $G_{\sigma}$ and $\Gamma_{\psi\psi\sigma}$.

In the first iteration of the vertex function one substitutes into 
the propagator equation the
tree-level value of the 3-point vertex which yields the requirement:
\begin{equation}
\frac{\delta\lambda_2^{(1)}}{3}v^2+\frac{\lambda_{3,R}^2}{18}v^2\left
(-i\int_kG_{\psi}(p-k)G_{\psi}(k)\right)_{div}=0.
\end{equation} 
The same sort of first iteration is done also in the integral equation of 
$\Gamma_{\psi\psi\sigma}$ and the two together lead to
\begin{equation}
\delta\lambda_2^{(1)}=-\frac{\lambda_{3,R}^2}{6}I_{\psi,div}, \qquad
\delta\lambda_3^{(1)}=-\frac{\lambda_{1,R}\lambda_{3,R}}{6}I_{\psi,div}.
\end{equation}
With this step also the finite parts are determined, which are
substituted into the second iteration. This step produces on the
right hand side of the expression $iG_{\psi}^{-1}$ the following
divergent piece in the full expression, which has to 
vanish:
\begin{equation}
\left[\frac{\delta\lambda_2^{(2)}}{3}+2\frac{\lambda_{3,R}\delta
\lambda_3^{(1)}}{18}I_{\psi,div}+\frac{\lambda_{3,R}}{9}
I_{\psi,F}(p)\left(\delta\lambda_3^{(1)}+\frac{\lambda_{1,R}\lambda_{3,R}}{6}
I_{\psi,div}\right)+\frac{\lambda_{1,R}\lambda_{3,R}^2}{108}
I^2_{\psi,div}\right]v^2=0.
\end{equation}
It is important that the subdivergence proportional to the finite
(possibly temperature or density dependent) $I_{\psi,F}$
automatically vanishes in view of the previous iteration step. This
feature of subdivergence cancellation 
is generally valid for each iteration step. The remaining
divergent terms determine the second 
term in the infinite series of $\delta\lambda_2$:
\begin{equation}
\delta\lambda_2^{(2)}=\frac{\lambda_{3,R}^2\lambda_{1,R}}{36}I^2_{\psi,div}.
\end{equation}
The second iteration of the equation of $\Gamma_{\psi\psi\sigma}$ determines
$\delta\lambda_3^{(2)}$ and 
$\delta\lambda_1^{(1)}=-\lambda_{1,R}^2I_{\psi,div}/6$. With simple
induction the following general recursions are found for the 
subsequent iterative terms of the counterterm series:
\begin{equation}
\delta\lambda_1^{(n)}=-\delta\lambda_1^{(n-1)}\frac{\lambda_{1,R}}{6}
I_{\psi,div},\quad
\delta\lambda_2^{(n)}=-\delta\lambda_3^{(n-1)}\frac{\lambda_{3,R}}{6}
I_{\psi,div},\quad
\delta\lambda_3^{(n)}=-\delta\lambda_1^{(n-1)}
\frac{\lambda_{3,R}}{6}I_{\psi,div},
\label{CounterRecur}
\end{equation}
where $n\ge1$ and $\delta\lambda^{(0)}_1:=\lambda_{1,R}$, 
$\delta\lambda_3^{(0)}:=\lambda_{3,R}$. \\
These equations imply the following non-perturbative renormalisation formulae:
\begin{equation}
\frac{1}{\lambda_1}=\frac{1}{\lambda_{1,R}}+\frac{1}{6}I_{\psi,div}, 
\qquad\frac{\lambda_3}{\lambda_1}=
\frac{\lambda_{3,R}}{\lambda_{1,R}},\qquad
\lambda_2-\frac{\lambda_3^2}{\lambda_1}=
\lambda_{2,R}-\frac{\lambda_{3,R}^2}{\lambda_{1,R}}.
\end{equation}
In addition to these relations between bare and renormalised couplings
we also obtain the renormalised sigma self-energy which is nothing
but the expansion of the last term in (\ref{sigmaMM})
in powers of $\lambda_1$ up to the  given order
of the iteration (of course, the couplings and the bubble integral 
have to be replaced by their renormalised expressions). 

It is easy to check that provided we would have known {\it a priori} 
these relations, they indeed ensure the simple form of the 
renormalised expression of (\ref{sigmaMM}) announced at the end of the 
previous section. The mass of the Higgs particle is determined by the 
pole of equation (\ref{sigmaMM}) ($iG^{-1}(p^2=M_\sigma^2)$=0) 
\begin{equation}
M_\sigma^2=\frac{v^2}{3}\left[\left(\lambda_{2,R}-
\frac{\lambda_{3,R}^2}{\lambda_{1,R}}\right)+
\frac{\lambda_{3,R}^2}{\lambda_{1,R}^2}\,\, 
\frac{1}{\lambda_{1,R}^{-1}-I_{\psi,F}(p^2=M_\sigma^2)/6}\right].
\label{Msigma}
\end{equation}
The renormalised quartic couplings are used also in the iterative
renormalisation of the equations for the order parameter $v$ and of
$G_{\psi}$ in which the tadpole is calculated with a self-consistent 
propagator. The mass counterterms $\delta m_2^2$ 
and $\delta m_3^2$ are determined as:
\begin{eqnarray}
&
\delta
m_2^{2(n)}=-\frac{\delta\lambda_1^{(n-1)}}{6}\frac{\Lambda^2}{16\pi^2}+
m_{2,R}^2\frac{\delta\lambda_3^{(n)}}{\lambda_{3,R}}=
-\frac{\delta\lambda_1^{(n-1)}}{6}\frac{\Lambda^2}{16\pi^2}+
m_{2,R}^2\frac{\delta\lambda_1^{(n)}}{\lambda_{1,R}},\nonumber\\
&
\delta
m_3^{2(n)}=-\frac{\delta\lambda_3^{(n-1)}}{6}\frac{\Lambda^2}{16\pi^2}+
m_{2,R}^2\frac{\delta\lambda_2^{(n)}}{\lambda_{3,R}}=
-\frac{\delta\lambda_3^{(n-1)}}{6}\frac{\Lambda^2}{16\pi^2}+
m_{2,R}^2\frac{\delta\lambda_1^{(n)}\lambda_{3,R}}{\lambda_{1,R}^2}
\label{MassRenorm}
\end{eqnarray}
implying
\begin{equation}
\frac{m_2^2}{\lambda_1}+\frac{\Lambda^2}{96\pi^2}=\frac{m_{2,R}^2}
{\lambda_{1,R}},\qquad
m_3^2-m_2^2\frac{\lambda_3}{\lambda_1}=m_{3,R}^2-m_{2,R}^2
\frac{\lambda_{3,R}}{\lambda_{1,R}}.
\end{equation}
The second equalities in both equations of (\ref{MassRenorm}) follow when 
one exploits (\ref{CounterRecur}).
Making use of these renormalised relations in (\ref{extrema}) and
(\ref{psiM}) one easily derives for the renormalised expression
of the vacuum expectation value the following formula:
\begin{equation}
\frac{1}{6}v^2=\left[
-m_{3,R}^2+\left(m_{2,R}^2-m_\psi^2\right)
\frac{\lambda_{3,R}}{\lambda_{1,R}}\right] 
\left(\lambda_{2,R}-\frac{\lambda_{3,R}^2}{\lambda_{1,R}}\right)^{-1}.
\label{vacuumvev}
\end{equation}

\section{Discussion}

The expressions for the Higgs mass (\ref{Msigma}) and the Higgs
condensate (\ref{vacuumvev}) show that the influence of the phantom
sector might be quite strong on both. In order to make the discussion
more transparent below we shall choose
\begin{equation}
\lambda_{2,R}=\lambda_{3,R}\equiv\lambda,\qquad
\lambda_{1,R}\equiv\lambda+\lambda'.
\end{equation}
In this case the gap equation (\ref{Eq:gap-equation}) which determines 
$m_\psi$ reads after the elimination of $v^2$ as
\begin{equation}
m_\psi^2=m_{2,R}^2-m_{3,R}^2+\frac{\lambda'}{6}T_{\psi,F},\qquad
T_{\psi,F}=\frac{m_\psi^2}{16\pi^2}\ln\left(\frac{e m_\psi^2}{M_0^2}\right).
\label{Eq:gap-dis}
\end{equation}
which has the approximate solution (for small $\lambda'$)
\begin{equation}
m_\psi^2\approx\frac{m_{2,R}^2-m_{3,R}^2}{1-\frac{\lambda'}{96\pi^2}
\ln\frac{e (m_{2,R}^2-m_{3,R}^2)}{M_0^2}}.
\end{equation}
For sufficiently large renormalisation scale $M_0$ one has
$0< m_\psi^2 < m_{2,R}^2-m_{3,R}^2$.
The Higgs condensate has the simple equation
\begin{equation}
\frac{\lambda}{6}v^2=-m_{3,R}^2+\frac{\lambda}{\lambda'}
(m_{2,R}^2-m_{3,R}^2-m_\psi^2).
\label{Eq:vev-dis}
\end{equation}
Depending on the ratio of the two quartic couplings and the
differences in the renormalised mass parameters we see that the
phantom fluctuations might produce a stable $Z_2$ symmetry breaking
condensate. This mechanism is similar to the dynamical symmetry
breaking of Coleman and Weinberg \cite{coleman77}.
The role of the electromagnetic field in their example is played
here by $\psi_i$, $\lambda'$ and the dynamically determined $m_\psi^2$.
Also the Higgs mass 
\begin{equation}
M_\sigma^2=\frac{v^2}{3}\frac{\lambda}{\lambda+\lambda'}
\left[\lambda'+\frac{\lambda}
{1-\frac{\lambda+\lambda'}{6}I_{\psi,F}(p^2=M_\sigma^2)}\right]
\label{Eq:sigma-dis}
\end{equation}
depends in a substantial way on both couplings which would give more 
flexibility in Higgs phenomenology.

In conclusion, we have presented an $O(N)$ symmetric model of the
phantom world interacting with the SM Higgs field, which was solved in
the $N=\infty$ limit. The solution obtained with iterative
renormalisation represents a non-trivial generalisation of the
well--known example of the pure $O(N)$ model. The method can be
applied to a large variety of models containing several multiplets 
with large number of components.

The existence of symmetry breaking solutions stabilised by the phantom
sector fluctuations was pointed out. In this respect we emphasise that
Eqs. (\ref{Eq:gap-dis}), (\ref{Eq:vev-dis}) and (\ref{Eq:sigma-dis})
are valid also at finite temperature, if $I_{\psi,F}$ and $T_{\psi,F}$
are evaluated accordingly. They can serve for investigating the nature
of finite temperature symmetry restoration in the generalised Higgs
sector.

Non-perturbative (lattice) studies at finite $N$ could clarify if our
findings are generically valid also for finite number of phantom field
components.

\section*{Acknowledgements}
We thank the participants of a Croatian-Hungarian Bilateral Workshop
in Theoretical Physics for valuable discussions on the first version of
this material.  Work supported by Hungarian Scientific Research Fund
(OTKA) under contract number T046129. Zs. Sz. is supported by OTKA
Postdoctoral Grant no. PD 050015. Support by the Hungarian Research
and Technological Innovation Fund, and the Croatian Ministry of
Science, Education and Sports is gratefully acknowledged. The authors 
acknowledge valuable comments of the Referee on the paper.

\end{document}